\documentclass[twocolumn,preprintnumbers,amsmath,amssymb]{revtex4-1}
\usepackage{graphicx}
\usepackage{dcolumn}
\usepackage{bm}
\usepackage[usenames,dvipsnames]{xcolor}

\begin{document}
\preprint{ \color{NavyBlue}  \today}

\title{\Large \color{NavyBlue}  Compound droplet manipulations on fiber arrays}

\author{F.Weyer\footnote{Correspondance : GRASP, Physics Department, University of Li\`ege, B-4000 Li\`ege, Belgium. http://www.grasp-lab.org }}
\author{M.Lismont}
\author{L.Dreesen}
\author{N.Vandewalle}

\maketitle
\setlength\parindent{0em}


{\bf 
Recent works demonstrated that fiber arrays may constitue the basis of an open digital microfluidics. Various processes, such as droplet motion, fragmentation, trapping, release, mixing and encapsulation, may be achieved on fiber arrays. However, handling a large number of tiny droplets resulting from the mixing of several liquid components is still a challenge for developing microreactors, smart sensors or microemulsifying drugs. Here, we show that the manipulation of tiny droplets onto fiber networks allows for creating compound droplets with a high complexity level. Moreover, this cost-effective and flexible method may also be implemented with optical fibers in order to develop fluorescence-based biosensor.
}

\vskip 0.2 cm

\section{Introduction}

\begin{figure}[h]
\begin{center}
\includegraphics[width=0.45\textwidth]{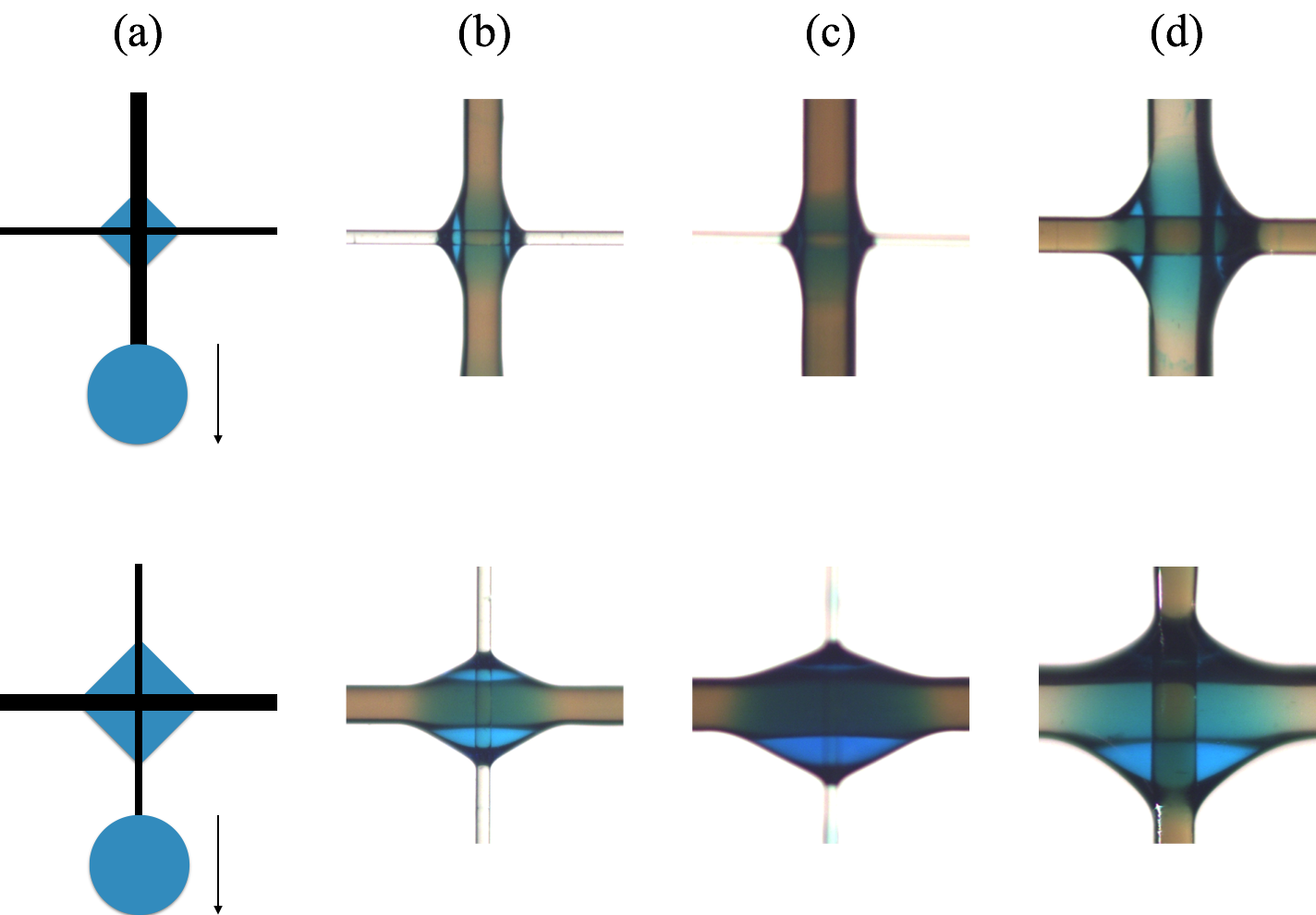}
\caption{Sketchs and pictures of tiny residual droplets remaining at an asymmetric fiber crossings after the motion of a large droplet ($8 \, {\rm \mu l}$) of colored water along the vertical nylon fiber. Four different cases are illustrated : (a) a schematic representation of a node made of two fibers with different diameters, (b) a water residue created at a crossing between a  $80 \, {\rm \mu m}$ diameter fiber and a $250 \, {\rm \mu m}$ diameter fiber, (c) a water residue created at a crossing between a  $80 \, {\rm \mu m}$ diameter fiber and a $350 \, {\rm \mu m}$ diameter fiber, and (d) a water residue created at a crossing between a  $250 \, {\rm \mu m}$ diameter fiber and a $450 \, {\rm \mu m}$ diameter fiber. Obviously, the volume of the residue depends on the fiber diameters. For each type of node, two configurations are possible : (top) the thicker fiber is vertical or (bottom) the thicker fiber is horizontal. The residues are quite different for these two situations : when the thick fiber is horizontal, the remaining droplet is seen to be much larger.}
\label{residu}
\end{center}
\end{figure}

The study of static droplets on fibers has been extensively reported regarding their geometry, their wetting properties and their lifetime \cite {mac1,mac3,lorenceau,sauret,mugele, eral,dufour,duprat}. However, Gilet \textit{et al.} demonstrated that droplets are also able to slide along fibers owing to the competition between gravity and capillary force \cite{giletb}. In the case of crossed fibers, defining a fiber node, droplet motion studies have shown that, depending on the volume of the droplets, $V$, and the fiber diameter, they might either remain pinned at the node or continue their way by passing trough the node, leaving behind tiny liquid amounts, as sketched in Figure \ref{residu}a. As shown in our previous work, this start-stop droplet motion along a fiber network may be used to separate an incoming droplet into several parts \cite{giletb}. Additionally to the droplet fragmentation, fiber networks may also be used to cause the coalescence or the chemical reaction between droplets \cite{gileta}. Another recent work proved that optical fibers network might be the starting point of optofluidic devices, allowing for probing the droplet content by spectrophotometry. Despite droplet motions concern a wide range of applications, droplet evaporation constitutes the major drawback of these innovative techniques. A way to circumvent this limitation is based on the encapsulation of the water core droplet into an oil shell \cite{weyer}, leading to the formation of a so-called \textit{compound droplet}.

Currently, the development of compound droplets, made of an oil drop containing several inner aqueous droplets, are still challenging considering scientific and technological applications involving drugs transport or hazardous materials manipulation. Although techniques are already implemented to create complex droplets \cite{planchette, planchette2, adams, weitz, utada, chu, weitzbis}, we intend, by using fiber arrays, to develop an easy way to generate and handle such droplets as well as to explore the physics behind. Moreover, this original approach allows for better controlling both the droplet size and the number of components inside the oil droplet. 

Our results are decomposed in three major successive steps. First, we show how to create the inner droplets through the droplet fragmentation at the network nodes. Second, we explain how tiny droplets can be released from the nodes using encapsulation by oil. Both colored and fluorescent inner droplets were considered. Finally, we combine fragmentation and release processes to create complex compound droplets, also called multiple microemulsions. 

\vskip 0.2 cm

\section{Experimental}
In order to achieve our tasks, we used a rigid frame in which nylon fibers or optical fibers are horizontally and vertically stretched to create a fiber network with a set of crossings. Since fiber diameter drives droplet motions, we investigate networks of fibers with diameters ranging from $80 \, {\rm \mu m}$ to $450 \, {\rm \mu m}$. Prior to experiments, nylon fibers were cleaned with acetone and distilled water before drying. Before their use, the multimode optical fibers with a diameter of $400 \, {\rm \mu m}$ are prepared as follow. From their distal end, 5 cm of the external jacket was removed with a fiber buffer stripping tool. The uncovered regions of fibers were cleaned with acetone to remove the polymer cladding and were rinsed with distilled water.

Droplets are formed using a mixture of SDS in water (0.01  ${\rm M}$) in which dye is added for getting stronger contrast. When optical fibers are used, fluorophores are dissolved inside the water solution instead of dye. Fluorescein and rhodamine were considered. The SDS molecules reduce the water surface tension and thus the contact angle of the droplet and so allow for generating barrel shaped droplet, which consists in a rotational symmetric pearl-like structure. Water droplets are left on the network and slide along the fibers. The water droplets pinned at the nodes are encapsulated by oil. We used a silicone oil (Dow Corning) with a density of $950  \, {\rm kg/m^3}$. Oil viscosity can affect the microemulsion lifetime so we investigate the encapsulation of water droplets with several viscosities ranging from $20 \, {\rm cSt}$ to $100 \, {\rm cSt}$. Oil and liquid mixtures are characterized by low surface tension and are therefore wetting the fibers. The surface tensions have been determined using the pendant drop method using a CAM 200 goniometer from KSV Instruments Ltd. The air-colored liquid surface tension is found to be $\gamma_{l} =   25.7 \pm 0.1 \, {\rm mN/m}$. The air-oil tension is $\gamma_{o} =   18.5 \pm 0.3 \, {\rm mN/m}$. The oil-colored liquid tension is much smaller : $\gamma_{ol} = 5.5  \pm 0.1 \, {\rm mN/m}$. 


\vskip 0.2 cm

\section{Results and discussion}

{\color{NavyBlue} \bf Droplet fragmentation} ---  In this section, we focus on the interaction between a water droplet and a node, which leads to the formation of a residue, a tiny droplet that will be encapsulated in an oil drop (see next section). When a single liquid droplet glides on a vertical fiber of diameter $b$, the droplet motion can be stopped by a horizontal fiber of diameter $a$. If capillarity overcomes gravity, the droplet sticks at the node. This occurs for a droplet volume smaller than a critical volume, $V_{stop}$, given by
\begin{equation}
V_{stop}= {\xi \gamma_{l} a \over \rho g}
\label{eq_stop}
\end{equation}
where $\rho$ is the liquid density, $g$ is the gravity acceleration and $\xi \approx 2 \pi$ is a shape factor depending mainly on the node geometry as shown by Gilet \textit{et al.} \cite{gileta}. The fiber diameter $a$ can be selected in order to trap droplets at selected sites on the array. For example, typical aqueous solutions on 200 $\mu$m nylon fibers imply $V_{stop} \approx 5 \, {\rm \mu \ell}$. If the incoming volume is larger then $V_{stop}$, the droplet slows down at the crossing, wets both fibers and detaches from the node leaving a tiny residual volume. The remaining volume, $V_r$, is much smaller than $V_{stop}$ and is expected to be strongly dependent on the geometrical parameters of the node, mainly the fiber diameters, but also on physical characteristics of the liquid. Pictures of Figure \ref{residu} show the residual droplets created after the crossing of a $8 \, {\rm \mu \ell}$ water droplet at different fiber nodes. Asymmetric nodes, \textit{i.e.} made of two fibers with different diameters ($a \ne b$), are presented. For each crossing, top row presents the case for which the thicker fiber is vertical and the thinner one is horizontal while the bottom row presents the same systems rotated by $90^\circ$. In this figure, (a) is a sketch of the crossings and (b), (c), and (d) are three nodes with different fiber diameters. The equilibrium shape is non-trivial but looks like a diamond. It has to be noticed that, for nodes with two different fibers, not only the diameter matters but also the position of these fibers. Indeed, the situation is completely different when the thick fiber is either vertical ($a<b$) or horizontal ($a>b$). It appears that a larger residual droplet remains at the node for a thick horizontal fiber. The typical volume of such droplets is ranging from $1 \, {\rm \mu \ell}$ on thick fibers down to $0.01 \, {\rm \mu \ell}$ for thin fibers. The latter volume is close to the one produced in closed microfluidic systems composed of channels \cite{weitz,utada,chu,adams,guz,weitzbis}. The manipulation of such tiny droplets on fibers opens therefore an alternative way to design microfluidic systems. 

The volume $V_{r}$ of the residual droplets has been measured for a wide range of nodes : not only different diameters but also different combinations of thin and thick fibers have been considered. The data are obtained by estimating the volume remaining at each node type from image analysis, as described in Supplementary Information. Figure \ref{compound}a shows the residual volume ($V_{r}$) evolution as a function of the horizontal diameter $a$ in the case of a vertical fiber with a diameter $b$ of $250 \, {\rm \mu m}$. This plot highlights the strong dependence between the residual volume and the diameter, $a$, of the horizontal fiber. As $a$ increases, $V_r$ becomes larger. For the other values of $b$, the evolution of $V_{r}$ is similar. The results are robust and seem independent of the volume of the incoming water droplet. 

\begin{figure}[h]
\begin{center}
\includegraphics[width=0.45\textwidth]{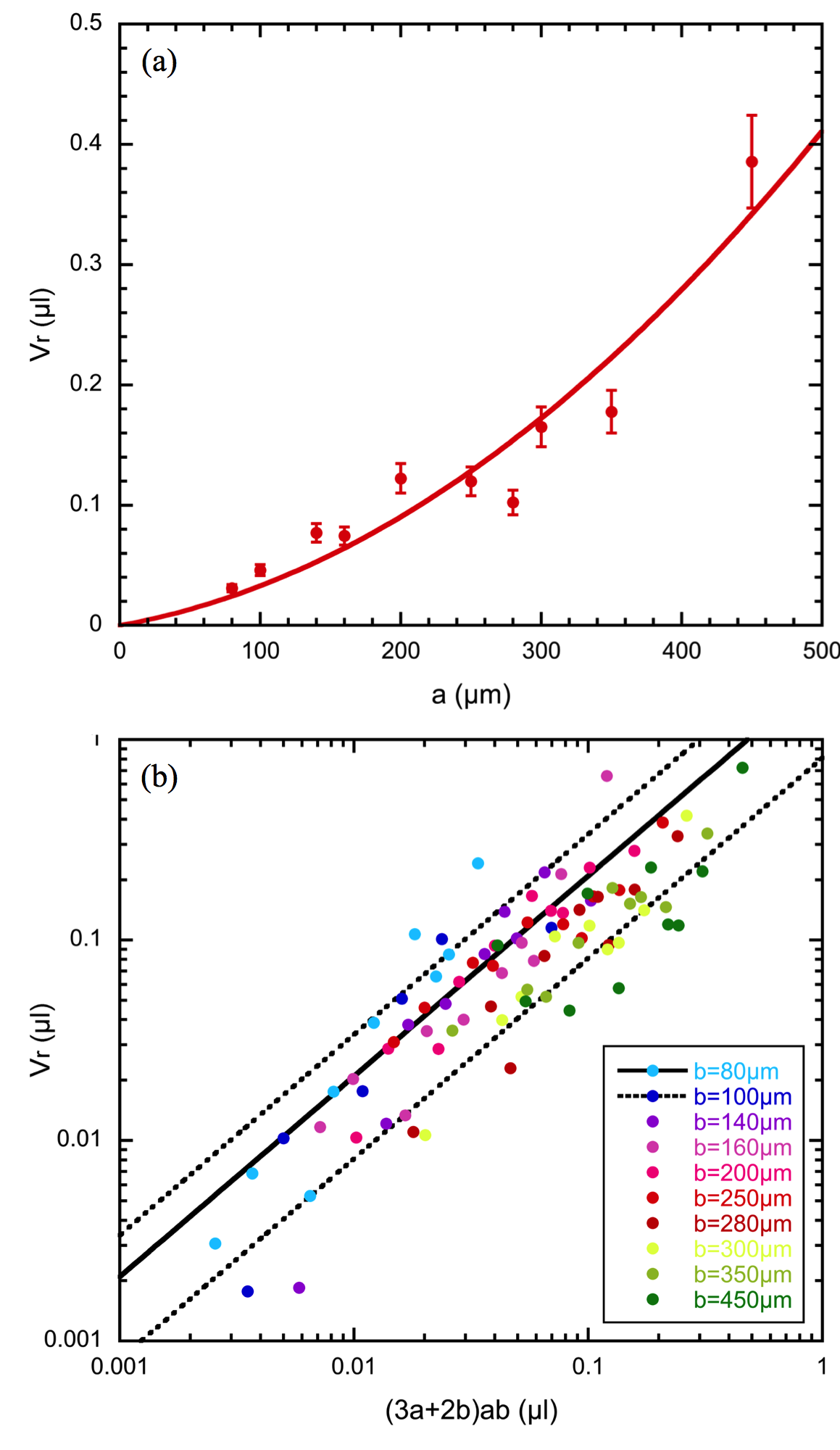}
\caption{(a) Plot of the residual volume, $V_r$, at a fiber crossing, as estimated by image analysis, as a function of the horizontal fiber diameter $a$. Data are shown for a fixed vertical diameter $b=250 \, {\rm \mu m}$. The curve corresponds to the model proposed by Eq. (\ref{eq_vr}), using only one single free fitting parameter $\zeta$. (b) The residual volume as a function of the suggested model shown for various vertical fiber diameters corresponding to the different colors. The plot highlights the linear dependance between the volume and the model. Moreover, the log-log plot emphasizes the wide range of volumes, i.e. over two decades, remaining on the fiber nodes. For each value of $b$, a specific value of $\zeta$ can be found and a linear fit can be plotted. We decided to show the linear fits corresponding to the average shape factor $\zeta = 2.09$ and to $\zeta = 2.09 \pm 1.28$ in dotted lines.}
\label{compound}
\end{center}
\end{figure}

We consider that the droplet fragmentation is entirely determined by the deformation of the droplet at the fiber crossing after being strongly slowed down, meaning that the capillary effects dominate the viscous ones and so that the fragmentation in independent of the incoming droplet speed. In order to confirm this assertion, we studied the droplet speed by analyzing movies of the incoming droplet motion. The typical droplet speed is found to be close to $0.01 \, {\rm m/s}$ at the node. The capillary number $Ca$, corresponding to the ratio between the viscous forces and the capillary forces, is calculated and is found to be equal to $0.06$. As $Ca$ is much smaller than $1$, viscous effects can be neglected. The incoming droplet will experience a deformation with a characteristic length $a$ as the wetting is directly related to the fiber diameters. Since the residual droplet has a diamond shape, we assume in a first approximation that it is made of four right triangles with a basis $a$ and a height $b$, corresponding to the horizontal and vertical fiber diameters, respectively. The rotation of this shape around the horizontal fiber using the Pappus-Guldinus theorem gives the approximate residual volume 
\begin{equation}
V_{r}= \zeta (3a+2b) ab
\label{eq_vr}
\end{equation}
where $\zeta$ is a shape factor, which is sensitive to the physical properties of liquids. In order to validate our model, we plot in Figure \ref{compound}b the measured residual volume as a function of $(3a+2b) ab$. The data show not only that $V_r$ depends linearly on $(3a+2b) ab$ but also that it is affected by any change of $b$ although this fiber is considered as a guide for the incoming droplet motion. Indeed, for each value of $b$, the scaling law is respected but, as $b$ increases, data are shifted towards larger values of $(3a+2b) ab$. Therefore, we assume that $\zeta$ depends on $b$ and the shape factor is calculated thanks to linear fits for every value of $b$. The shape factor $\zeta$ varies from $0.99$ to $4.97$ with an average of $2.09$ and a standard deviation of $1.28$. The fits corresponding to  $\zeta = 2.09$ and $2.09 \pm 1.28$ are plotted in Figure \ref{compound}b. These observations show that the scaling proposed in our model is coherent with a shape factor depending on $b$. The model curve for $b=250 \, {\rm \mu m}$ is shown by the red solid lines in Figure \ref{compound}a. The agreement between the model and the data is satisfactory on both plots. Moreover, an asymmetry between horizontal and vertical fibers is visible on the plot : when switching the roles between $a$ and $b$, the volume of the residual droplet is quite different. Please note that our model is consistent with this observation : the volumes are different when swapping $a$ and $b$ in Eq.(\ref{eq_vr}), i.e. when the system is rotated by $90^\circ$.

{\color{NavyBlue} \bf Releasing microdroplets} --- 
After having studied the fragmentation of a large droplet and having characterized the residual volume, the detachment and the collection of the residue is investigated in order to create a compound droplet. The fragmentation process of a large droplet into several tiny volumes can be implemented on a vertical array of fibers. The method allows to control both position and volume of each tiny droplet. However, the release of a specific residue could be seen as a non-trivial task. Indeed, once formed, the residual volume $V_r$ is trapped at the node because it is much smaller than $V_{stop}$ and therefore the capillary forces hold the residue at the crossing whether the fibers are rotated or agitated. We investigate the droplet release by launching an oil droplet. This oil drop will encapsulate the residue and will release it in some specific conditions.

\begin{figure}[h]
\begin{center}
\includegraphics[width=0.45\textwidth]{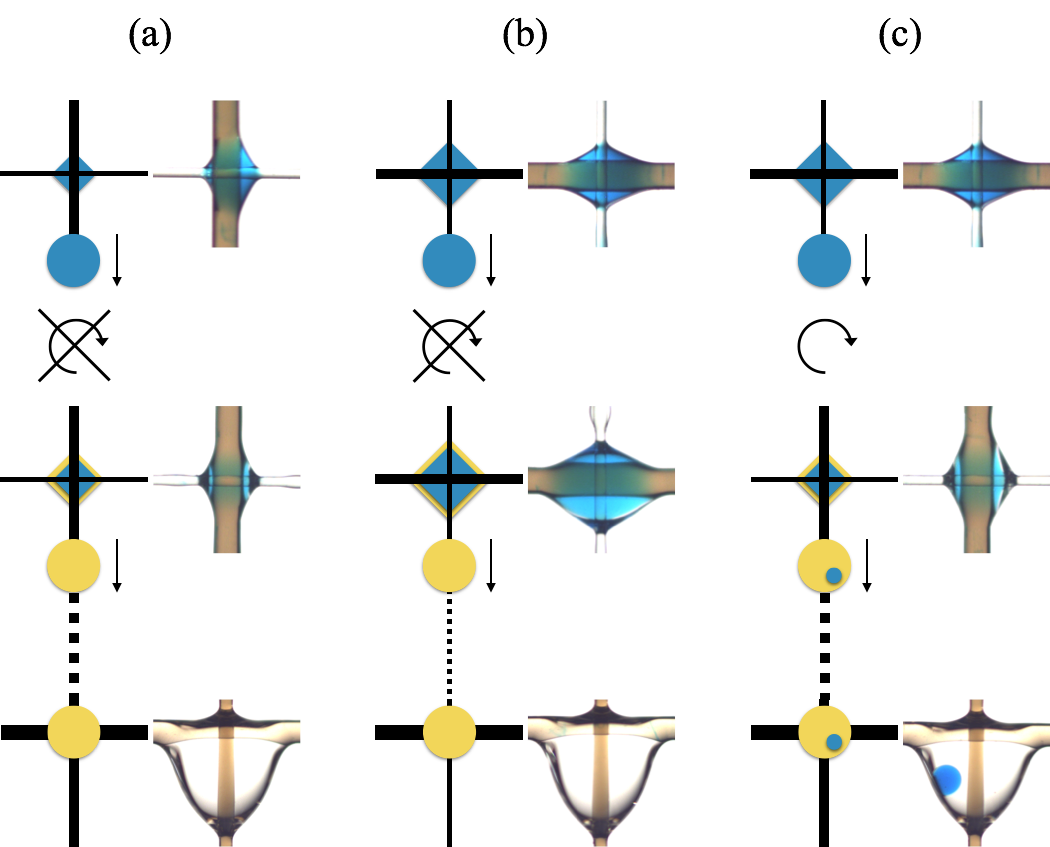}
\caption{Sketch of the creation and the collection of the residues. For the three cases, first (top row), a large water droplet crosses a node and leaves a residue. Second, an oil drop glides along the fiber and crosses the node (middle row). Finally (bottom row) the oil droplet continues its way and is stopped by a larger horizontal fiber. In the first two cases, the thicker fiber remains vertical (a) or horizontal (b) during the whole process. Therefore, once the residue is created at the node, it remains trapped and no water is released from the node. Whereas, in case (c), the thicker fiber is horizontal during the creation of the water residue and then, after the rotation of the system by $90^\circ$, it becomes vertical and, the residual volume is, now, larger than expected and therefore the oil droplet releases the exceeding volume of water.}
\label{proof}
\end{center}
\end{figure}

Once the water residue is created with a volume $V_r$, an oil droplet is launched on the vertical fiber and meets the residual water droplet on the node. Engulfing of the core aqueous phase by oil is ensured by the competition of air/oil/water surface tensions. The encapsulation takes place \cite{guz} if
\begin{equation}
\gamma_{l} > \gamma_{o} + \gamma_{ol}
\label{eq_encapsulation}
\end{equation}
which is verified in our case (see Experimental for the values of the surface tensions). Then, the instantaneously formed compound droplet detaches from the node when its global volume is larger than $V_{stop}$ and leaves a residue given by $V_r$ following the same process as described in the previous section. This residue is mainly made of water with a thin oil film. As the water volume corresponds to $V_r$, the aqueous droplet remains trapped at the node and the oil crosses without releasing any water. This process is sketched in Figure \ref{proof}a and b. The top row presents the aqueous residual droplets at the same node except that the fibers in (b) are rotated by $90^\circ$ with respect to (a), leading to a slightly larger volume $V_r$. However, in both cases, when an oil droplet glides on the vertical fiber, it just wraps the residue, leaves a thin oil layer and continues its way without collecting any water (middle row). Thus, at the end, the oil, stopped by a much larger fiber, does not contain water (bottom row). In order to release water residues, increasing the volume at the node by adding oil is not sufficient. Taking advantage of the asymmetry of the residues when $a$ and $b$ are switched is a solution. Indeed, on Figure \ref{proof}c, an aqueous residue, with a volume $V_r$ is created while the thicker fiber is horizontal (top row). The system is then rotated by $90^\circ$, so that the thicker fiber is vertical and the thinner one is horizontal. At this stage, the volume of the residue is larger than expected with Eq.(\ref{eq_vr}) because it was created while the thick fiber was horizontal. However, this volume still smaller than $V_{stop}$, thus it remains at the node. Therefore, when the oil droplet arrives at the node, the excess volume is dragged by the oil droplet and is encapsulated inside the moving droplet. A tiny residue is so trapped at the node with the volume expected for this fiber configuration (middle row).  At the end of the fiber array, the oil droplet is stopped by a crossing of two large fibers. Compared to cases (a) and (b), the oil droplets contains a tiny aqueous droplet. Assuming that the aqueous residue is always determined by Eq.(\ref{eq_vr}), the inner volume, $V_{in}$, released by the node inside the oil droplet is given by 
\begin{equation}
V_{in}= \zeta (a-b) ab
\label{eq_in}
\end{equation}
where $a$ and $b$ are again the horizontal and vertical fiber diameters before the rotation. Since the asymmetry drives the process, one is able to release volume around $V_{in} \approx 0.1 \, {\mu \ell}$. However, an optical effect, due to the oil shell surrounding the inner droplets, prevents the measurement of precise water volume by image analysis. Similar results have been obtained for different oil viscosities between 20 and 100 cSt but we focused on the 100 cSt oil because it ensures a longer lifetime for the inner droplets.

These results show that the encapsulation allows for releasing a residue trapped at a node. When the oil droplet reaches the node, the excess water volume detaches so that the oil droplet contains a water component. This process is similar to dripping such that the main relevant parameter is the geometry of the node. But more importantly, the rotation of $90^\circ$ provides an exchange of thick and thin fibers such that the release is triggered.

\begin{figure}[h]
\begin{center}
\includegraphics[width=0.45\textwidth]{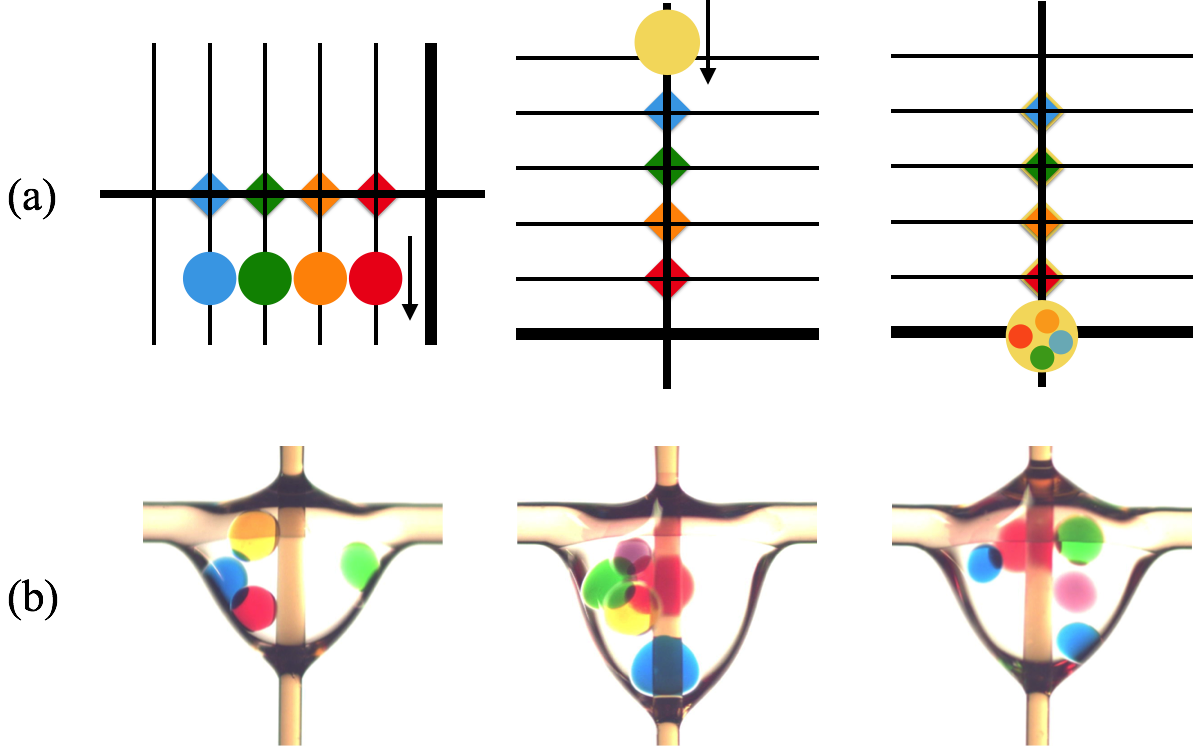}
\caption{(a) Sketch of the basic principle to create compound droplets. We use an array of fibers of different diameters. First, four water droplets of different colors slide along the vertical fibers, cross the nodes and leave a residue. Second, the system is rotated and an oil drop glides along the new vertical fiber and collects a part of each water residue. In the end, a compound droplet is obtained. (b) Picture of microemulsions obtained by using the device sketched in (a). By changing the number of fibers, the number of inner droplets can be increased.}
\label{principle}
\end{center}
\end{figure}

{\color{NavyBlue} \bf Complex compound droplets} ---The last step in creating multicompound droplets is to encapsulate more than one inner droplet. By repeating the release of tiny droplets, we can increase the number of inner droplets and therefore create a microemulsion. Here, we prove that using fibers is also a suitable way to generate droplets with multiple inner components. The principle for creating multicomponent droplets is sketched on Figure \ref{principle}a. Here, four vertical fibers are used to create four residues on a single horizontal fiber. As the droplets are wetting the fibers and are large enough, they cross the node and they each leave a residue. The remaining tiny droplets are on a common fiber which is then placed vertically, by rotating the whole fiber array by $90^\circ$. A silicon oil droplet is then released from top to bottom. This droplet collects a proportion of every droplet and form a compound system. A larger fiber at the bottom of the array stop the motion of the silicon oil droplet, allowing for analysis or further delivery. A single row of fiber nodes is therefore needed for creating a compound droplet. The first picture in Figure \ref{principle}b illustrates the generation of a microemulsion containing four different inner components. The four different colors inside the microemulsion prove that the oil drop has collected a part of every water droplet. The number of inner components can be increased as shown in the second and the third pictures of Figure \ref{principle}b with five and six inner droplets. These pictures prove that our technique is efficient at generating microemulsions. A 2D/3D array could provide a promising way to multiplex or handle a much larger set of droplets. 

\begin{figure}[h]
\begin{center}
\includegraphics[width=0.45\textwidth]{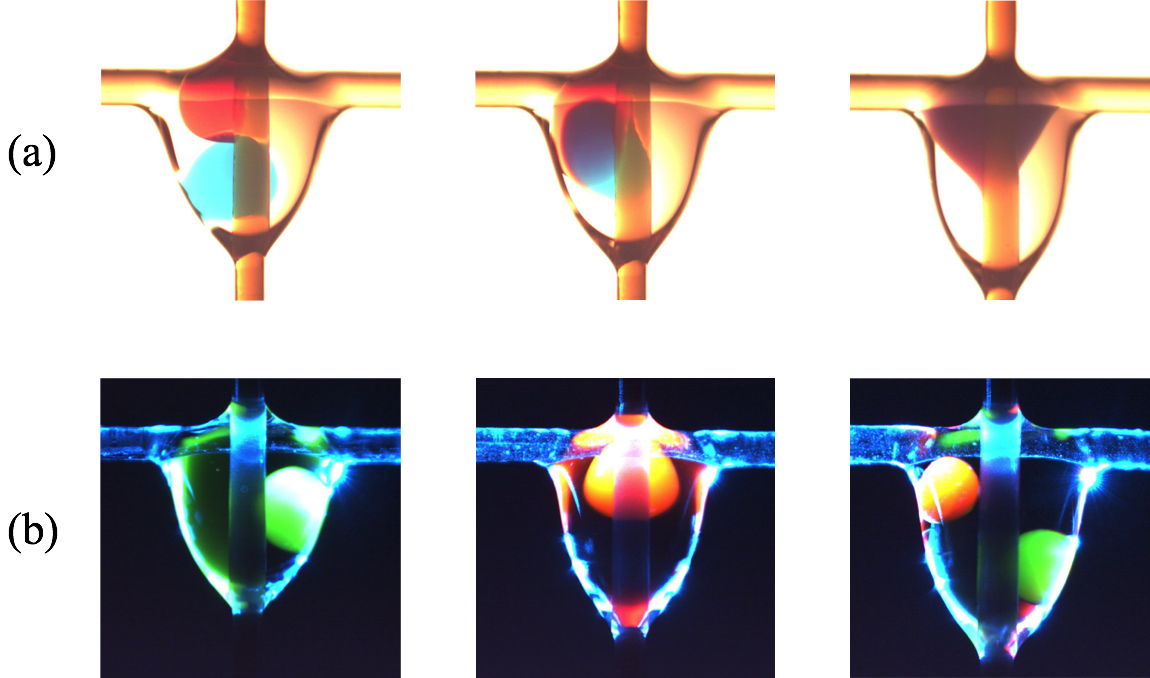}
\caption{Pictures of several applications of compound droplets. (a) The evolution of the inner components over time. Reagents can be placed inside the components and interact after the fusion of the droplets. This reaction is represented by the change in colors after the coalescence. (b) By replacing the horizontal fiber by an optical one, fluorescence of inner droplets can be excited. Fluorescein and rhodamine are excited by a blue laser and produce a green and orange fluorescence, respectively.}
\label{micro}
\end{center}
\end{figure}


\vskip 0.2 cm

\section{Conclusion}

The fabrication of a complex droplet by encapsulating tiny drops has an enormous potential for scientific and technological applications in pharmaceutical drug delivery, biosensors, soft materials, microreactors, food science, cosmetics, etc. Microfluidic devices have been invented for producing some elaborated fluidic objects, also called « multiple microemulsions » or « compound droplets ». However, the on-demand production of such drops, their manipulation and their further analysis are a real challenge. We proved herein that fiber arrays become the central point of a new type of microfluidic device able to produce highly sophisticated droplets, as illustrated in Figure \ref{principle}b. We first showed how a droplet interacts with a crossing leading to the formation of residues. Second, we proposed a technique for collecting those water droplets inside an oil drop. And finally, we obtained microemulsions by using this process. The device is very effective because the number and the volume of the inner drops can be controlled by selecting the fiber diameters. Placing different pure components on different fibers avoids the contact between them and allows them to remain separate inside the oil drop once they are collected. Moreover, the coalescence of the inner droplets can be triggered by using a specific paraffin (Suppocire AIM oil \cite{adams}) instead of oil. This multicomponent droplet production offers a number of possibilities. Indeed, arrays can be designed for generating, on a single device, all possible combinations of liquid mixtures starting from few drops (see Supplementary Information). Once created, several complex droplets can also be manipulated on the same array. As shown in Figure \ref{micro}a, reactions can also take place inside a compound droplet. Reagents, placed in the inner components, interact together after the fusion of the inner droplets. This type of reaction is represented by the mixture of the colors after the coalescence of the drops. Moreover, when encapsulated drops are engineered for being microreactors or smart sensors, the fibers themselves could be used to excite fluorescence or to measure an optical signal from the inner drops. Figure \ref{micro} (b) shows that it is possible to excite and detect the fluorescence of the components by using optical fibers coupled with a laser. A 488 ${\rm nm}$ laser beam is guided into a multimode fiber through a fiber coupler. The incident light propagates along the fiber and interacts with the fluorophores dissolved inside the inner droplets. The fluorescence light is then collected by the other optical fiber and is conveyed to a CCD spectrometer. This technique has already been validated in our lab \cite{lismont}. However, in this work, it is applied to the water droplets inside the oil shell. Fibers provide therefore an original way to create and manipulate compound droplets. This effective and adaptable method makes possible the applications requiring highly sophisticated droplets.



\end{document}